# Growth of Diamond Films from Tequila

J. Morales[1,2], L.M. Apátiga[2] and V.M. Castaño[2,*]

1.-Facultad de Ciencias Físico Matemáticas

Universidad Autónoma de Nuevo León

Av. Universidad S/N, San Nicolás, Nuevo León, México 66450

ippajmc@yahoo.com.mx

2. Centro de Física Aplicada y Tecnología Avanzada

Universidad Nacional Autónoma de México,

Boulevard Juriquilla 3001, Santiago de Queretaro, Querétaro, México 76230

apatiga@servidor.unam.mx

castano@fata.unam.mx

*corresponding author

**Abstract**

Diamond thin films were growth using Tequila as precursor by Pulsed Liquid Injection Chemical Vapor Deposition (PLI-CVD) onto both silicon (100) and stainless steel 304 at 850 °C. The diamond films were characterized by Scanning Electron Microscopy (SEM) and Raman spectroscopy. The spherical crystallites (100 to 400 nm) show the characteristic 1332 cm$^{-1}$ Raman band of diamond.

**Introduction**

The Pulsed Liquid Injection Chemical Vapor Deposition (PLICVD) technique has been employed to growth diamond thin films using a mixture of acetone and water as precursor [1-3]. As in any CVD technique, the Pulsed Liquid Injection Chemical Vapor Deposition involves many reactions in vapor phase as well as surface processes. In this case, vapor-phase reaction is driven by the flash evaporation phenomena to produce reactive products at the reaction chamber. In general, the PLI-CVD offers many advantages, such as uniform deposition over large area, conformal coverage, selective deposition and high reproducibility [4-7].

Many hydrocarbon, fluorocarbon, and other organic sources have been employed to produce diamond films [1-8]. Typically, the promotion of diamond bonding over graphitic bonding is only accomplished when the percentage of hydrocarbon in the gas phase is relatively small. For a number of technical reasons, efforts to change the diamond process from a molecular-hydrogen based process have been made in the last years. Also, O- and OH-containing chemicals have been also explored in both plasma-assisted and hot-filament techniques. For example, small quantities of oxygen and water vapor have been added to microwave plasma reactors for the purpose of oxygen addition. Small percentages (0.5 to 2%) of oxygen and small percentages of water vapor (0 to 6%) improve the quality of the Raman spectra and decrease the deposition temperature. However, higher percentages of both have been demonstrated to degrade the diamond quality. Therefore, a careful C-O-

H relationship is a must if pure diamond phase is to be successfully produced in a CVD-type system.

On the other hand, Tequila is a wide-known alcoholic beverage, granted origin denomination since 1982, eight year are necessary to grow and cultivated this agave.  When the agave plant from which tequila is produced is ready to be processed, it is cooked with vapor and under pressure and the juice is extracted, fermented and distilled twice to obtain a solution with 55 % of alcohol content. Then, the alcoholic solution is diluted with distilled water to obtain a final product (38 to 43 % alcohol content) and finally, aged in different containers, depending on the tequila kind desired.  As we shall see in what follows, tequila, or at least some types of it, present naturally the adequate atomic composition to achieve a proper diamond nucleation.

**Experimental**

Small pieces of a Si (100) wafer and commercial stainless steel (type 304) were used as substrates, fixed to the holder through silver paste. Temperature was controlled at 850 °C through an automatic PID temperature control (Eurotherm). Reactor pressure varied from 4.76 to 4.99 Torr due to the injection processes and to the flash evaporation phenomena. The carrier and reaction gases flux were fixed at 0.8 and 0.1 l/min, respective. "Tequila blanco" (white tequila) Orendain brand, a clear, un-aged liquor distilled from the juice of blue agave (Agave Tequilana) plant [9], was used as precursor. This tequila, 80 proof and with C-H-O atomic relationships of 0.37 C, 0.84 H and 0.29 O (Figure 1), was injected at a frequency of 2 pulses per second (500 ms) with an opening time of 4 ms. A total of 21768 pulses were applied in each experiment and a micro dose of 6.26 x $10^{-3}$ ml was injected per pulse (Table 1). Temperatures in the evaporation zone and along the vapor transport line were fixed at 280 °C. The deposit was studied through a Dilor micro-Raman spectrometer with a 20 mW, 632 nm He-Ne laser equipped with a confocal microscope and a JEOL Low-Vacuum Scanning Electron Microscope (JSM-6060LV), operating at 15 kV, secondary electrons, spot 50 and WD 11 mm.

**Results and discussion**

Tequila was injected at a rate of 6.26 x 10$^{-3}$ ml) per pulse from high to low pressure (4.5 psi to a few torrs) in an argon flow at 280 °C, prior to flash evaporation. Tequila basically consists of water and ethanol, both molecules will be dissociated through the reactor zone. Because of the bonding energies of C-C, C-H, C-O and H-OH, the ethanol molecule will be dissociated to supply the carbon atoms with hybrid bond (sp$^3$), whereas water provides an excess of hydrogen to form other allotropes.

The reactor temperature activates the silicon 3s and 3p orbitals to form the primitive partial covalent bond with the carbon hybrid orbital sp$^3$. As the temperature increases, the surface reactions rates and the mobility of the adsorbed species are enhanced, thus increasing the growth rate [11]. Additionally, the surface roughness of the silicon substrate favors the nucleation process [10].

The corresponding C-H-O ratios, represented in figure 1, show that Tequila lies within the Diamond Growth Region in the C-H-O diagram by Bachmann et al. [1]. Figures 2 and 3 show the Raman spectrum and SEM micrographs of diamond thin films deposited at 850 °C onto silicon and stainless steel 304 substrates, respectively.

The corresponding Raman spectra do not show the 1140 and 1490 cm$^{-1}$ modes, but only the diamond mode at 1332 cm$^{-1}$. The modes 1140 and 1490 cm$^{-1}$ appear in the nanocrystalline diamond thin films (NCD)   because of the formation of  trans-polyacetylene [12,13].  The SEM micrographs show micro crystallites from 100 to 500 nm in diameter.

## Conclusion

This work presents a new and original precursor (Tequila) to growth micro crystalline diamond thin films by the PLICVD technique. Nucleation and growth are strongly related to the nature of the Si and stainless steel substrates surface. Tequila has the adequate C-O-H atomic ration to grow Diamond Like Carbon thin films using the PLICVD technique. This technique is an excellent alternative to produce industrial-scale diamond thin films for practical applications using low cost precursors.

## Acknowledgements

The authors wish to thank M. en I.Q. Alicia del Real for the SEM studies and Dr. Roberto Sato, Dr. S. Jiménez and Ing. F. Melgarejo for kind access to their Raman facilities. The authors are also indebted to Dr. Victor Puntes for valuable discussions and comments. The technical assistance of Mr. A. Loeza is acknowledged, as well.

**Table 1**

**PLICVD parameters using Tequila as precursor.**

| PLICVD PARAMETERS | |
|---|---|
| Precursor | Tequila |
| Substrate Temperature | 850 °C |
| Reactor Pressure | 4.76 TO 4.99 TORR |
| Total Carrier Gas Flow | Ar, 1.0 L |
| Evaporation Temperature | 280 °C |
| Injector Frecuency | 500 ms |
| Pulse Length | 4 ms |
| Dosis | $6.26 \times 10^{-3}$ ml/Pulse |

**Figure Captions**

**Figure 1.** Atomic C-H-O diagram showing the diamond domain.

**Figure 2.** SEM micrograph and Raman spectrum from a diamond thin film deposited on a Si substrate.

**Figure 3.** SEM micrograph and Raman spectrum from diamond thin films growth on a Stainless steel substrate .

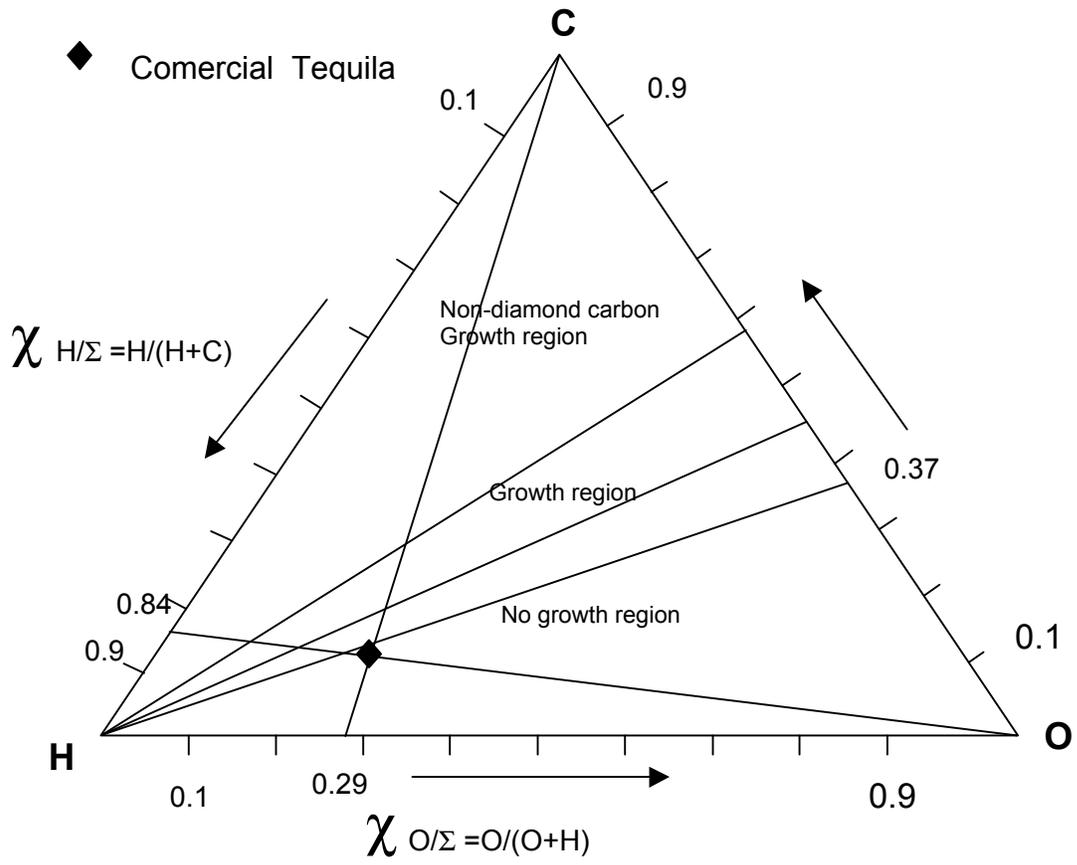

Figure 1

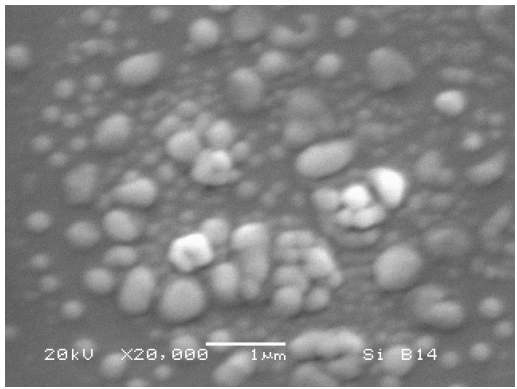 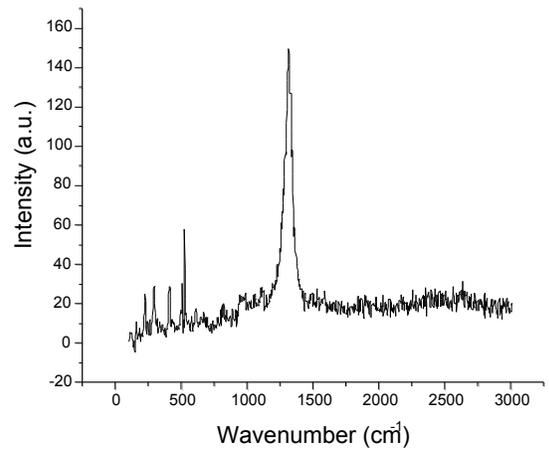

Figure 2.

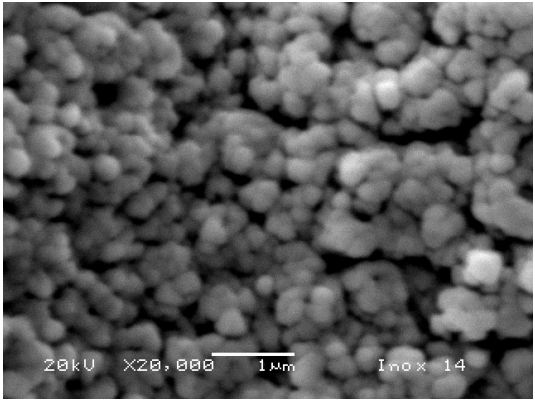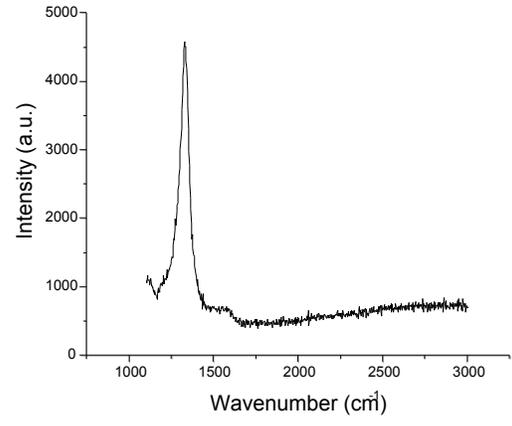

Figure 3.